\documentclass{article}
\usepackage{spconf,amsmath,graphicx}
\usepackage{bm}
\usepackage{pgfplots}
\usepackage{flushend}
\usepackage[ruled,vlined]{algorithm2e}
\usepackage{tikz}
\usepackage{hyperref}
\usepackage{textcomp}
\usepackage{xcolor}
\usepackage{nicefrac}
\usepackage[export]{adjustbox}
\usepackage{tabu,multirow,colortbl}
\usepackage{hhline}
\DeclareUnicodeCharacter{2212}{−}
\usepgfplotslibrary{groupplots,dateplot}
\usetikzlibrary{patterns,shapes,arrows,arrows.meta,chains,fit,positioning,calc}
\pgfplotsset{compat=newest}
\usetikzlibrary{positioning}
\newlength\figureheight
\newlength\figurewidth
\tikzstyle{box} = [draw]
\tikzstyle{boxs} = [draw,blur shadow={shadow blur steps=5},shadow xshift=0.1cm,shadow yshift=-0.03cm,fill=white]
\AtBeginShipout{\AtBeginShipoutUpperLeft{%
		\put(\dimexpr\paperwidth-20cm\relax,-0.5cm){\parbox[t]{19cm}{\copyright 2024 IEEE. Published in 2024 International Conference on Acoustics, Speech and Signal Processing (ICASSP), scheduled for 14-19 April 2024 in Seoul, Korea. Personal use of this material is permitted. However, permission to reprint/republish this material for advertising or promotional purposes or for creating new collective works for resale or redistribution to servers or lists, or to reuse any copyrighted component of this work in other works, must be obtained from the IEEE. DOI: \url{https://doi.org/10.1109/ICASSP48485.2024.10448350}}}%

}}
\title{Color Agnostic Cross-Spectral Disparity Estimation}
\name{Frank Sippel, Nils Genser, Hannah Och, Jürgen Seiler, and André Kaup}
\address{Friedrich-Alexander-Universität Erlangen-Nürnberg\\
	Multimedia Communications and Signal Processing\\
	Cauerstraße 7, 91058 Erlangen, Germany\\
\thanks{The authors gratefully acknowledge that this work has been supported by
the Deutsche Forschungsgemeinschaft (DFG, German Research Foundation) under project number 491814627. \\
\noindent\hspace*{5mm}%
Source code: \textit{\url{https://github.com/FAU-LMS/cade}}.
}}

\DeclareMathOperator*{\median}{median}
\newcommand{\mn}{m,\!n}

\newlength{\imgWidth}

\begin{document}
\ninept
\maketitle
\begin{abstract}
	Since camera modules become more and more affordable, multispectral camera arrays have found their way from special applications to the mass market, e.g., in automotive systems, smartphones, or drones.
    Due to multiple modalities, the registration of different viewpoints and the required cross-spectral disparity estimation is up to the present extremely challenging.
    To overcome this problem, we introduce a novel spectral image synthesis in combination with a color agnostic transform.
    Thus, any recently published stereo matching network can be turned to a cross-spectral disparity estimator.
    Our novel algorithm requires only RGB stereo data to train a cross-spectral disparity estimator and a generalization from artificial training data to camera-captured images is obtained.
    The theoretical examination of the novel color agnostic method is completed by an extensive evaluation compared to state of the art including self-recorded multispectral data and a reference implementation.
    The novel color agnostic disparity estimation improves cross-spectral as well as conventional color stereo matching by reducing the average end-point error by 41\,\% for cross-spectral and by 22\,\% for mono-modal content, respectively.
\end{abstract}
\begin{keywords}
Deep Learning, Disparity Estimation, Multispectral Imaging
\end{keywords}
\section{Introduction}
\label{sec:intro}
\begin{figure}[t]
	\centering
	\small
	\hspace*{-1em}
	\begin{tikzpicture}
		\node[below left] (img) at (-4.5,2.05) {\includegraphics[width=0.40\columnwidth]{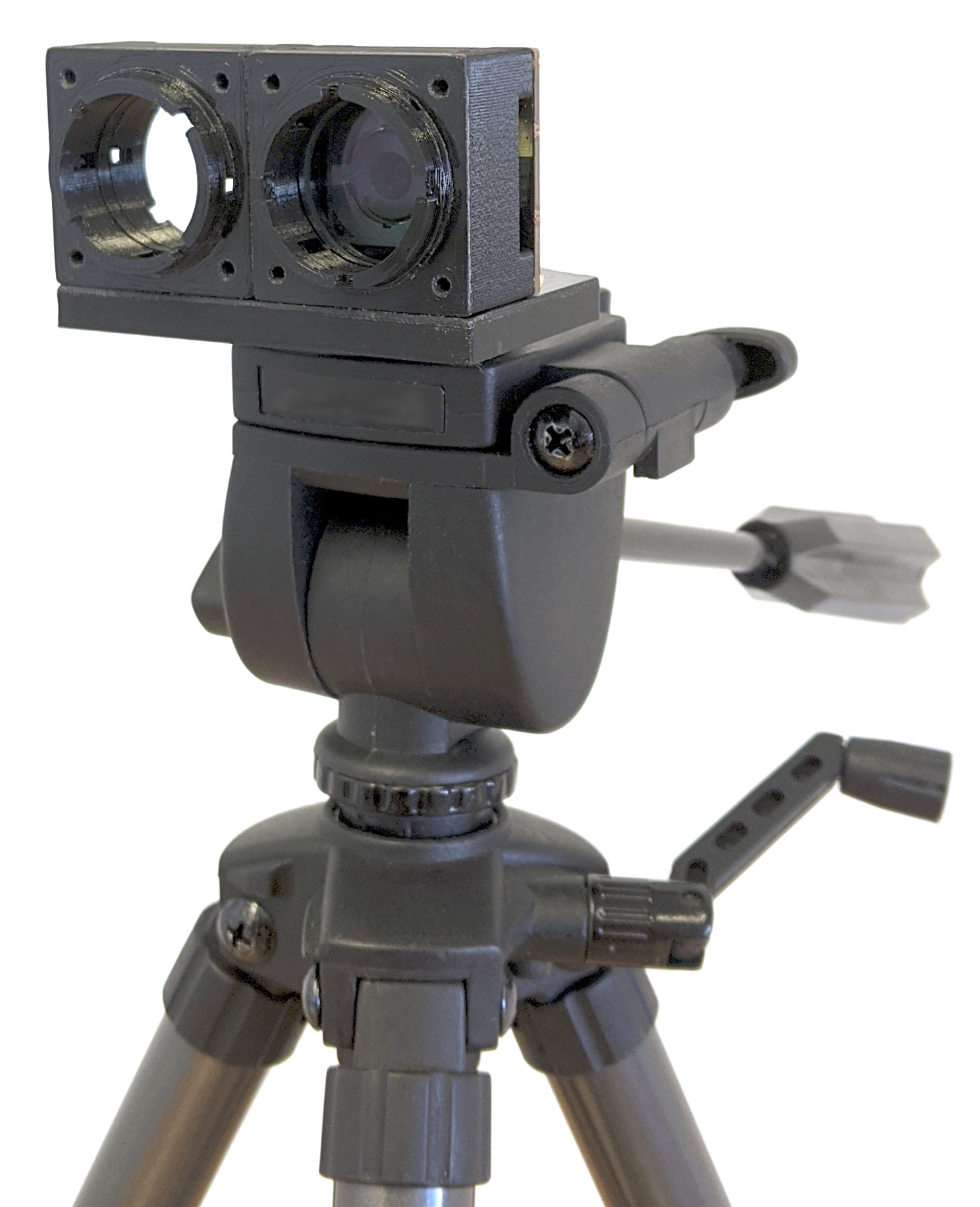}};
		\node at (-7.2,2.15) {Multispectral stereo camera};
		\node[below left] (img) at (-0.1,2.5) {\includegraphics[width=0.48\columnwidth]{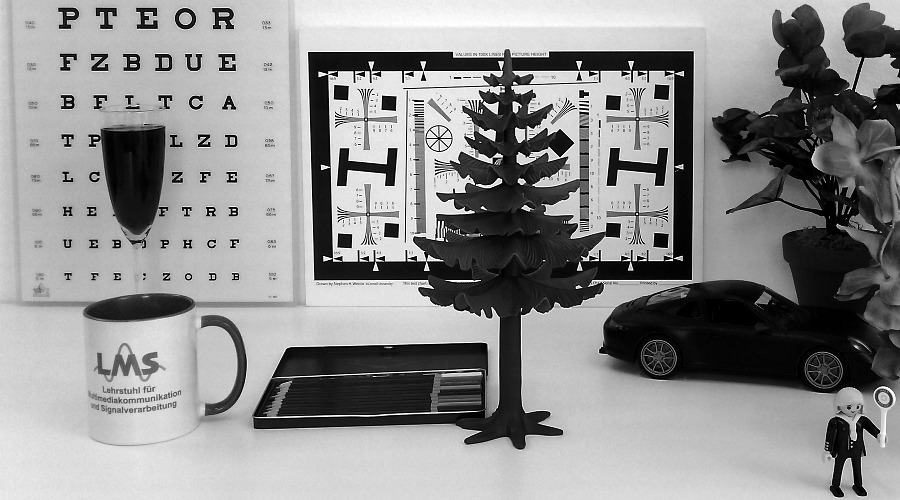}};
		\node[below left,fill=white,fill opacity=0.75,text opacity=1] at (-0.1,2.5) {\scriptsize Arbitrary filter, e.g., green \ };
		\node[below left] (img) at (-0.1,0) {\includegraphics[width=0.48\columnwidth]{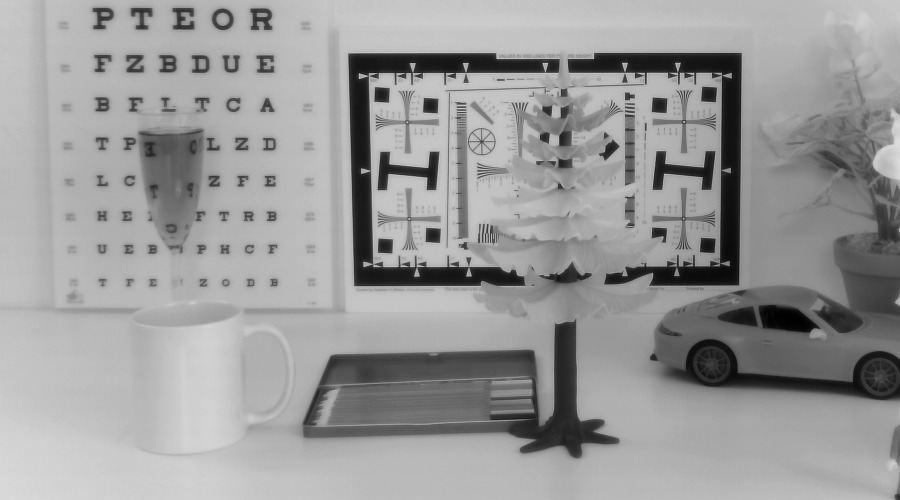}};
		\node[below left,fill=white,fill opacity=0.75,text opacity=1] at (-0.1,0) {\scriptsize Arbitrary filter, e.g., near-infrared \ };
	\end{tikzpicture}
	\vskip-1em
	\caption{On the left, a self-manufactured multispectral stereo camera is shown. The filters can be flexibly chosen by hand. The setup is employed to demonstrate the practical application of the proposed algorithm. On the right, two exemplary recordings are depicted.}
	\label{img:prototype}
	\vskip-1em
\end{figure}

In recent years, multi-camera setups are found in various electronic devices like drones~\cite{wierzbicki2018} or smartphones~\cite{lg2018} as image sensors are more and more affordable.
Instead of using similar cameras at different positions (mono-modal), it has become popular to measure multiple image modalities, e.g., by combining color (RGB) and near-infrared (NIR) cameras.
The Apple iPhone TrueDepth IR system~\cite{apple2018}, which combines a multi-lens NIR camera with an infrared emitter, is a well-known representative.
Once multispectral and depth information has been acquired, it can be used for plenty of applications, e.g., to collect vital signs~\cite{rapczynski2018}, or for biometric identification~\cite{kumar2012}.
However, these use cases require the measurement of accurate depth information, such that spectral images from different viewpoints can be registered to a common reference.
Apparently, this requires a high-quality cross-spectral stereo matcher.
We introduce a novel color agnostic disparity estimation, which significantly outperforms state-of-the-art methods and additionally is applicable to mono-modal stereo imaging.
Instead of proposing yet another stereo matcher, our novel algorithm is able to turn any stereo network into a cross-spectral disparity estimator.
To demonstrate the performance of our algorithm for practical applications, a self-manufactured prototype has been designed for recording test images as shown in Fig.~\ref{img:prototype}.

\section{State of the Art}
\label{sec:soa}
For cross-spectral stereo matching, approaches that combine a structural similarity cost function like Census transform~\cite{zeglazi2017} or Zero-Normalized Cross Correlation (ZNCC)~\cite{mattoccia2008} along with a cost aggregation such as semi-global matching (SGM)~\cite{hirschmuller2005} are employed in practical setups.
In addition to this strategy, \cite{kim2017DASC} adapts the correlation function according to the actual image content.
While classical approaches achieve a sufficient quality in multi-camera arrays~\cite{genser2020_CAMSI}, stereo setups generate noisy and inaccurate disparity maps.
Even though Convolutional Neural Networks (CNNs) have defined a new baseline for accurate color disparity estimation, they are hardly used in cross-spectral setups due to the lack of suitable multispectral training data.
To overcome the problem of missing training data, only few unsupervised deep learning strategies like~\cite{zhi2018_CSStereo,liang2019} have been proposed which achieve a similar performance as recent conventional stereo matchers.
Moreover, cross spectral networks explicitly for VIS-NIR setups~\cite{han_crossspectral_2023} have been developed, however, they fail to generalize to arbitrary spectral bands.

In contrast, CNNs already lead to breakthroughs in fast as well as high quality mono-modal stereo matching due to the availability of large amounts of training data.
Initial approaches like~\cite{zbontar2015} focused on computing matching distances while the final disparity was still calculated by applying the classical steps from literature: cost aggregation, disparity computation, and refinement~\cite{scharstein2001}.
Recent CNNs use an end-to-end learning architecture~\cite{Xu2023_CVPR, Xu_2022_CVPR,zhang2019GANet,yin2019,chang2018}, which is roughly divided in three subtasks.
First, spatial pyramid pooling layers set up a cost volume.
Second, the cost volume is aggregated via convolutional layers comparable to classical SGM. Third, a regression is conducted to obtain a disparity map~\cite{kendall2017}.

In~\cite{genser2020_CSDL}, it is shown that a simplistic training data generation can be successfully used to retrain mono-modal stereo matchers for cross-spectral disparity estimation.
In the following, we introduce a novel color agnostic method to significantly improve on state of the art.

\section{Color Agnostic Disparity Estimation}
\label{sec:CADE}
\begin{figure}[t]
	\vskip-0.2em
	\hspace*{-0.6em}
	\begin{tikzpicture}[>=latex']
		\small
		\input{./lib/tikzlibrarydsp}
		\setlength{\imgWidth}{0.2\columnwidth}

		\node[below right,box,minimum height=5.25cm,minimum width=9.0cm,dotted] (novelBox) at (0.02,-2.35) {};
		\node[below,align=center,fill=white] (novelLabel) at (4.5,-2.12) { \ Novel approach \ };

		\node[below,align=center] (label) at (2,-0.6) {RGB training images};
		\node[below right] (img) at (0,-0.95) {\includegraphics[viewport=270 157 470 270,clip,width=\imgWidth,frame]{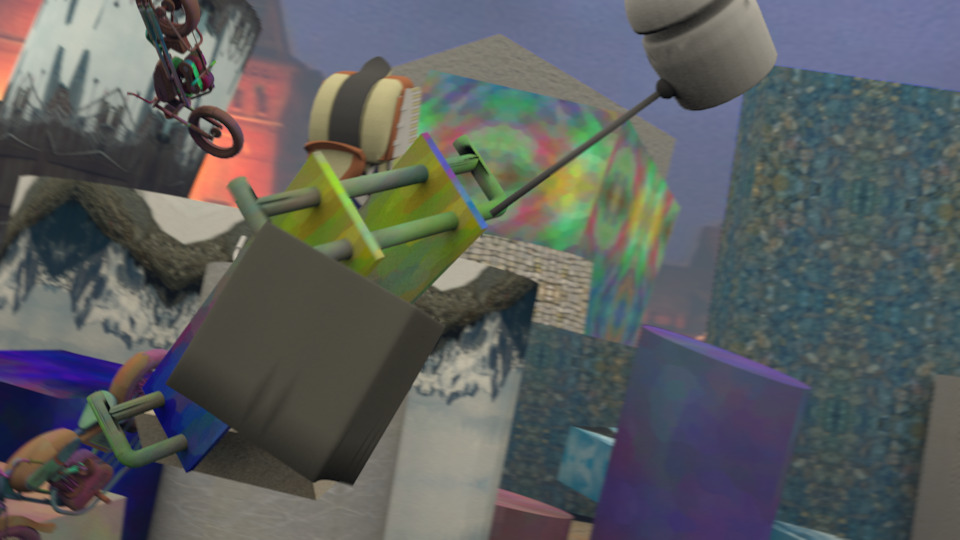}};
		\node[below right] (img) at (2,-0.95) {\includegraphics[viewport=270 157 470 270,clip,width=\imgWidth,frame]{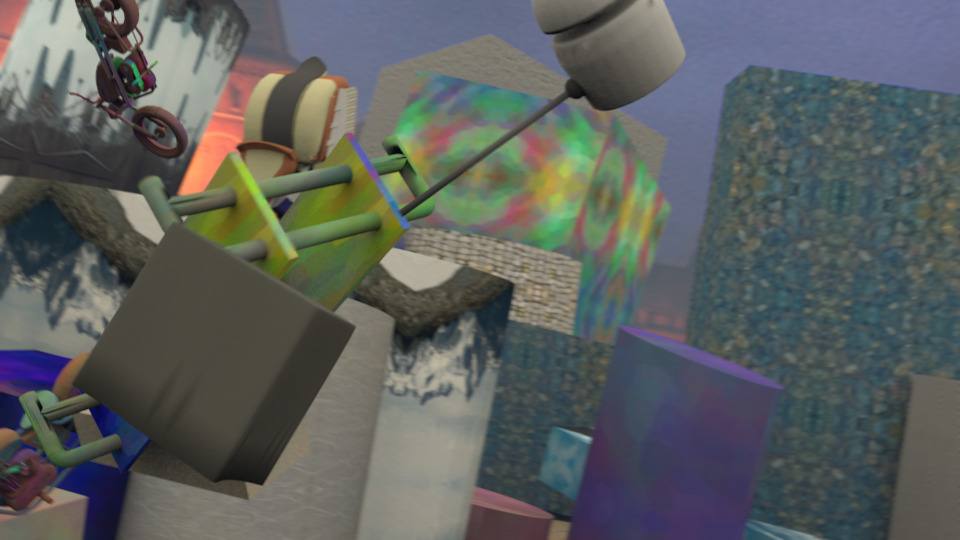}};
		\node (trainOOut) at (2,-2.1) {};
		\node (trainGInp) at (2,-2.6) {};
		\node[below right] (img) at (0,-2.75) {\includegraphics[viewport=270 157 470 270,clip,width=\imgWidth,frame]{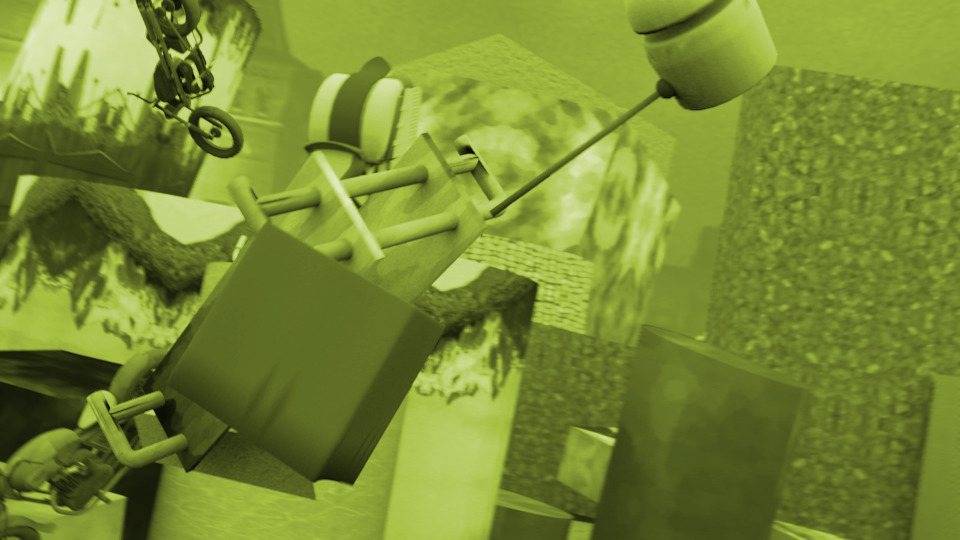}};
		\node[below right] (img) at (2,-2.75) {\includegraphics[viewport=270 157 470 270,clip,width=\imgWidth,frame]{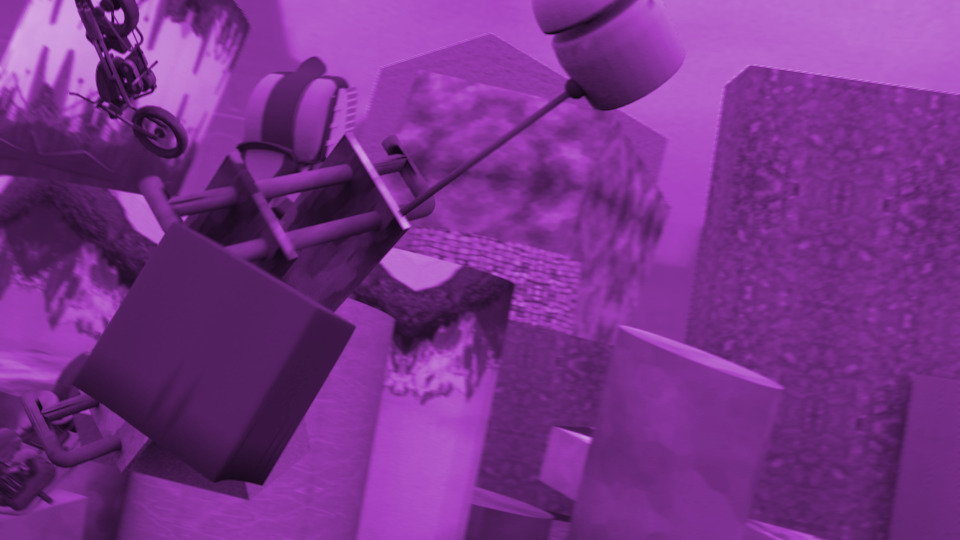}};
		\node (trainGOut) at (2,-3.9) {};
		\node (trainFInp) at (2,-4.4) {};
		\node[below right] (img) at (0,-4.55) {\includegraphics[viewport=270 157 470 270,clip,width=\imgWidth,frame]{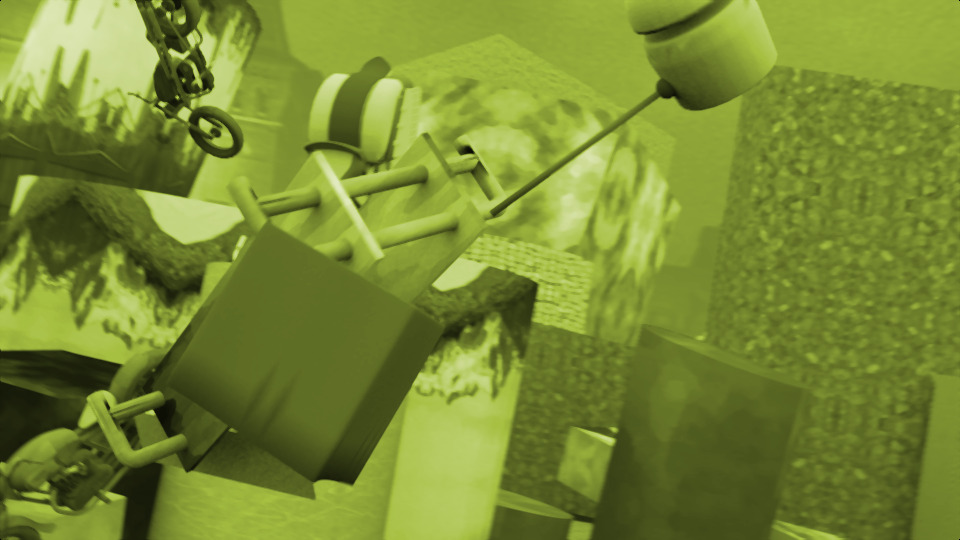}};
		\node[below right] (img) at (2,-4.55) {\includegraphics[viewport=270 157 470 270,clip,width=\imgWidth,frame]{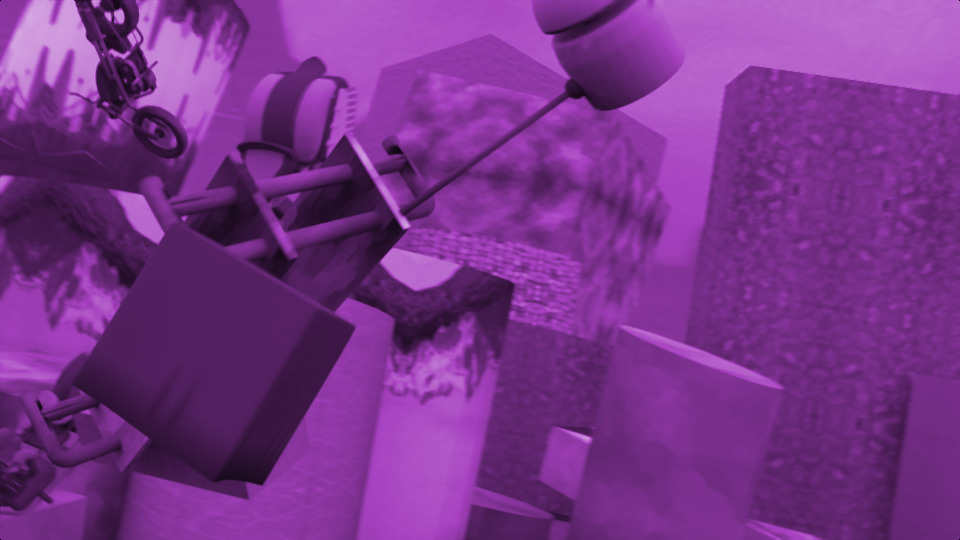}};
		\node (trainFOut) at (2,-5.7) {};
		\node (trainTInp) at (2,-6.2) {};
		\node[below right] (imgTL) at (0,-6.35) {\includegraphics[viewport=270 157 470 270,clip,width=\imgWidth,frame]{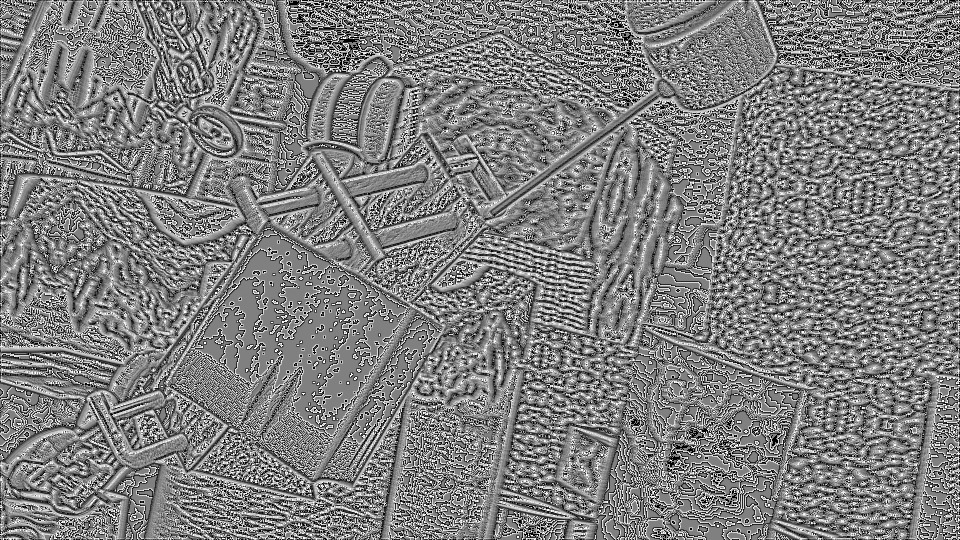}};
		\node[below right] (imgTR) at (2,-6.35) {\includegraphics[viewport=270 157 470 270,clip,width=\imgWidth,frame]{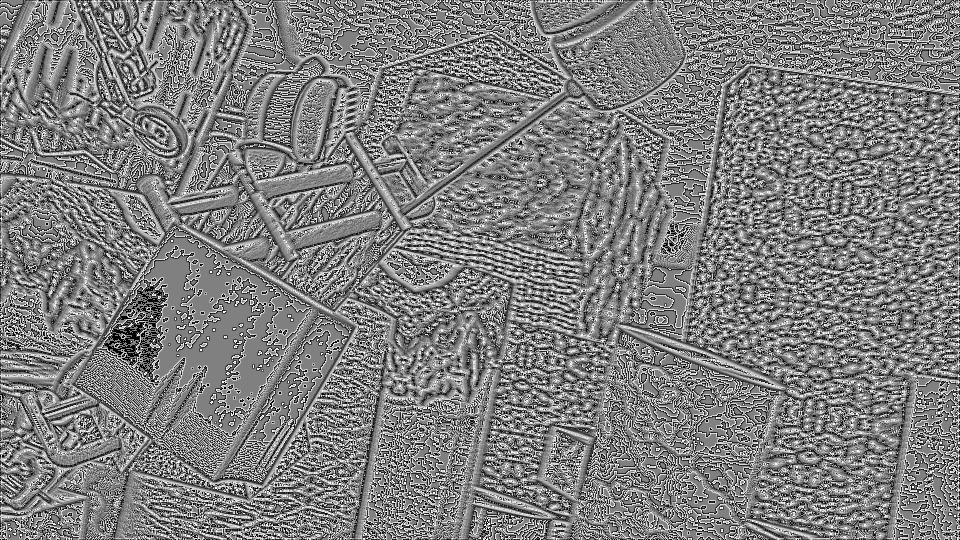}};
		\node[below right,box,minimum height=0.75cm] (cnnT) at (1.125,-7.85) {\begin{tabular}{c}Disparity\\estimator\end{tabular}};
		\node[below right] (trainDL) at (0,-9.0) {\includegraphics[viewport=270 157 470 270,clip,width=\imgWidth,frame]{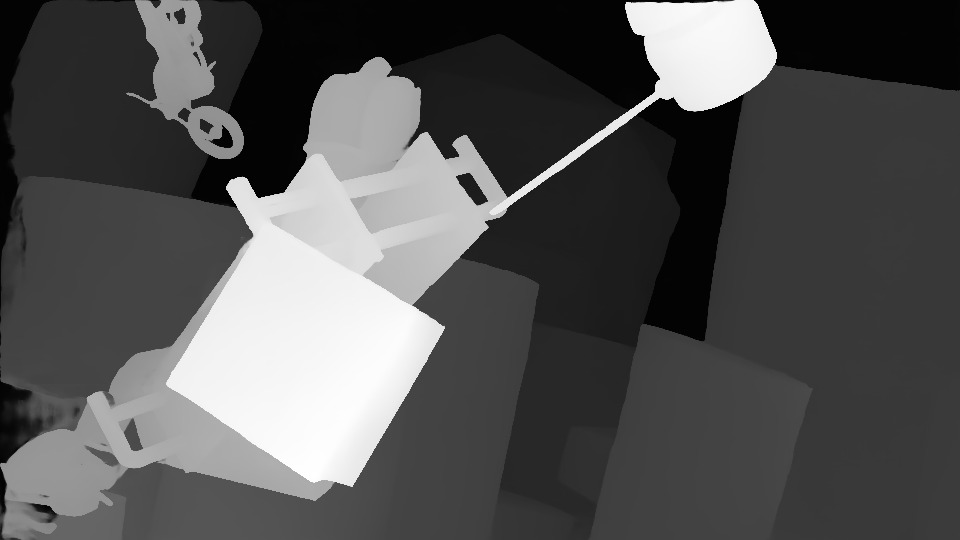}};
		\node[below right] (trainDR) at (2,-9.0) {\includegraphics[viewport=270 157 470 270,clip,width=\imgWidth,frame]{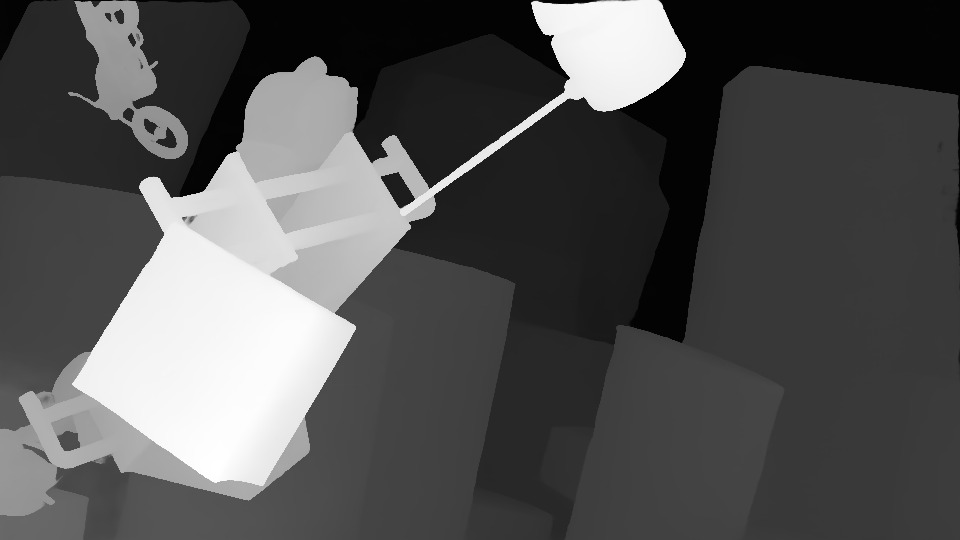}};
		\node[below right] (training) at (0,-10.1) {$\underbrace{\qquad\qquad\qquad\qquad\qquad\qquad}_{\text{\normalsize Training}}$};
		\draw[->,thick] (trainOOut) -- (trainGInp) node [pos=1.7] {\footnotesize Spectral band syn. (Sec.\,\ref{sec:CADE-syn})};
		\draw[->,thick] (trainGOut) -- (trainFInp) node [pos=1.7] {\footnotesize Denoising (Sec.\,\ref{sec:CADE-ca})};
		\draw[->,thick] (trainFOut) -- (trainTInp) node [pos=1.7] {\footnotesize \ Color agnostic model (Sec.\,\ref{sec:CADE-ca})};
		\draw[->,thick] (imgTL.south) -- (cnnT) node [pos=0.5] {};
		\draw[->,thick] (imgTR.south) -- (cnnT) node [pos=0.5] {};
		\draw[->,thick] (cnnT) -- (trainDL.north) node [pos=0.5] {};
		\draw[->,thick] (cnnT) -- (trainDR.north) node [pos=0.5] {};

		\node[below,align=center] (label) at (7,-0.6) {Multispectral recordings};
		\node[below right] (img) at (5,-0.95) {\includegraphics[viewport=270 157 470 270,clip,width=\imgWidth,frame]{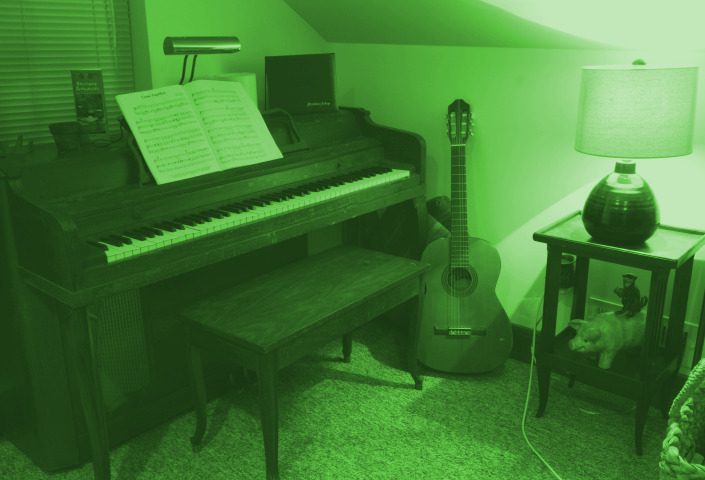}};
		\node[below right] (img) at (7,-0.95) {\includegraphics[viewport=270 157 470 270,clip,width=\imgWidth,frame]{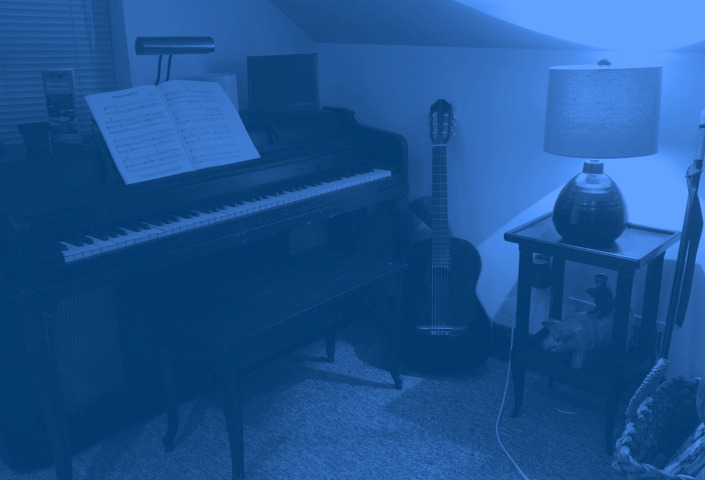}};
		\node (infeOOut) at (7,-2.1) {};
		\node (infeFInp) at (7,-4.4) {};
		\node[below right] (img) at (5,-4.55) {\includegraphics[viewport=270 157 470 270,clip,width=\imgWidth,frame]{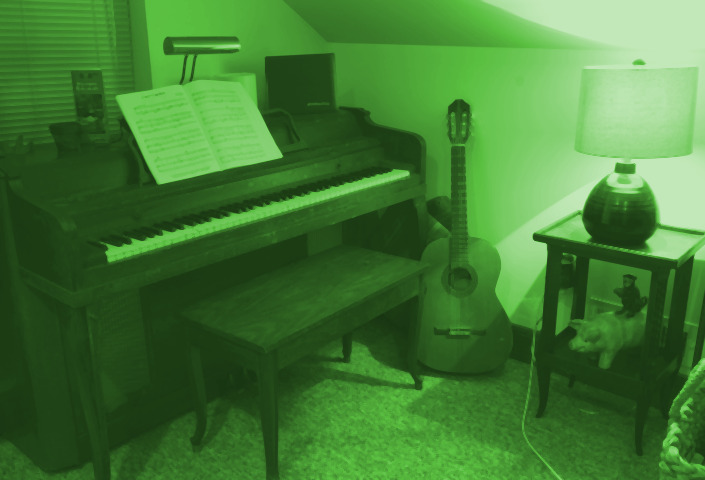}};
		\node[below right] (img) at (7,-4.55) {\includegraphics[viewport=270 157 470 270,clip,width=\imgWidth,frame]{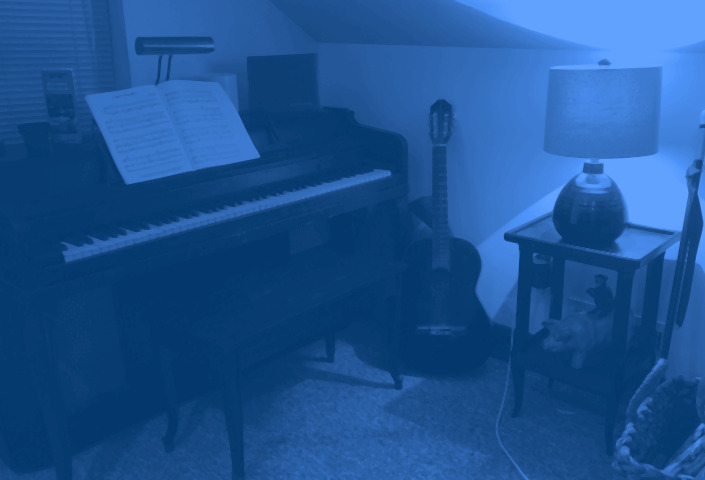}};
		\node (infeFOut) at (7,-5.7) {};
		\node (infeTInp) at (7,-6.2) {};
		\node[below right] (imgIL) at (5,-6.35) {\includegraphics[viewport=270 157 470 270,clip,width=\imgWidth,frame]{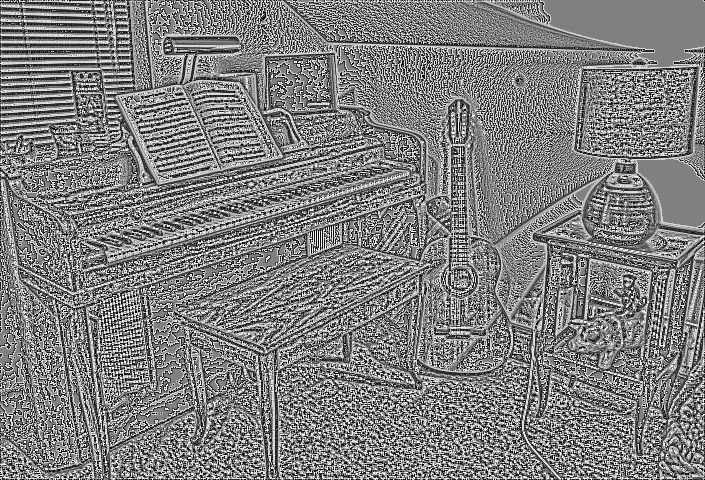}};
		\node[below right] (imgIR) at (7,-6.35) {\includegraphics[viewport=270 157 470 270,clip,width=\imgWidth,frame]{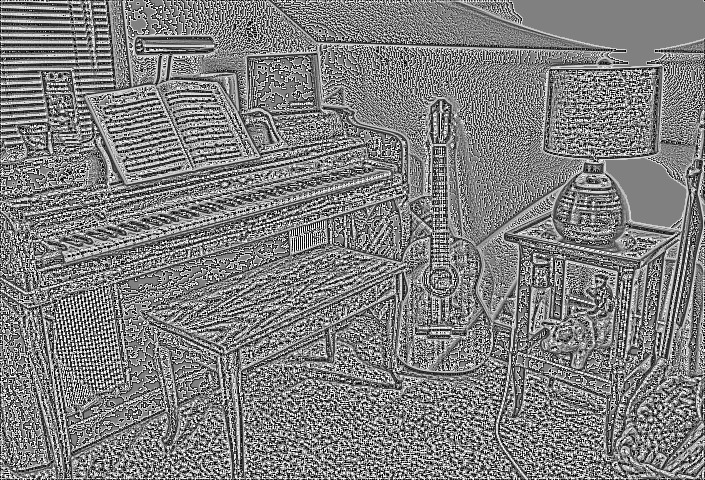}};
		\node[below right,box,minimum height=0.75cm] (cnnI) at (6.125,-7.85) {\begin{tabular}{c}Disparity\\estimator\end{tabular}};
		\node[below right] (infeDL) at (5,-9.0) {\includegraphics[viewport=270 157 470 270,clip,width=\imgWidth,frame]{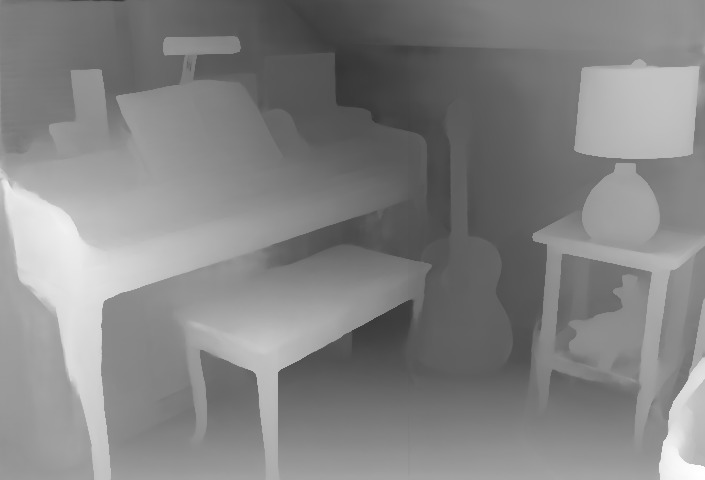}};\\
		\node[below right] (infeDR) at (7,-9.0) {\includegraphics[viewport=270 157 470 270,clip,width=\imgWidth,frame]{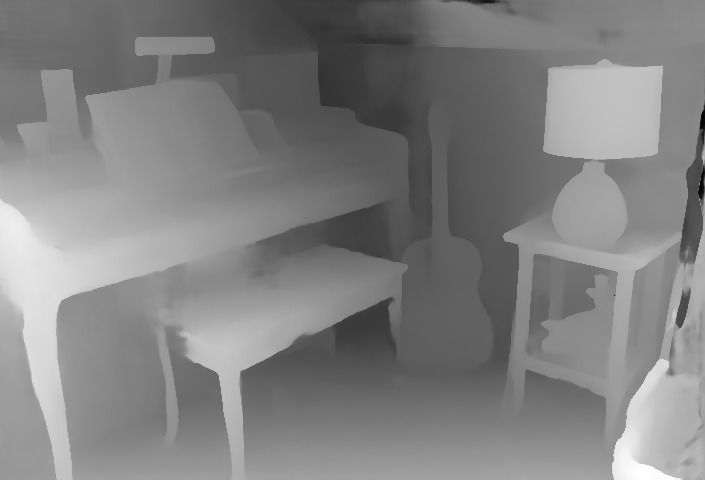}};
		\node[below right] (training) at (5,-10.1) {$\underbrace{\qquad\qquad\qquad\qquad\qquad\qquad}_{\text{\normalsize Inference}}$};
		\draw[->,thick] (infeOOut) -- (infeFInp) node [pos=1.1] {\footnotesize Denoising (Sec.\,\ref{sec:CADE-ca})};
		\draw[->,thick] (infeFOut) -- (infeTInp) node [pos=1.7] {\footnotesize Color agnostic model (Sec.\,\ref{sec:CADE-ca})};
		\draw[->,thick] (imgIL.south) -- (cnnI) node [pos=0.5] {};
		\draw[->,thick] (imgIR.south) -- (cnnI) node [pos=0.5] {};
		\draw[->,thick] (cnnI) -- (infeDL.north) node [pos=0.5] {};
		\draw[->,thick] (cnnI) -- (infeDR.north) node [pos=0.5] {};
		\draw[->,thick,dashed] (cnnT) -- (cnnI) node [pos=0.5,above] {\footnotesize Trained model} node [pos=0.5,below] {\footnotesize (Sec.\,\ref{sec:CADE-sum})}; 
	\end{tikzpicture}
	\vskip-1.1em
	\caption{Overview of the novel color agnostic disparity estimation. On the left, a disparity estimator is trained using widely available RGB stereo datasets and our novel color agnostic model. In the same way, camera-captured multispectral recordings are processed on the right using the trained model. The proposed color agnostic transform neglects the differences between artificial and camera-captured as well as RGB and multispectral data.}
	\label{img:CADE-intro}
	\vskip-1em
\end{figure}

The goal of CADE is to transform any stereo matching CNN into a cross-spectral disparity estimator.
For this, we introduce the following procedure.
First, a spectral image synthesis is presented which significantly extends our above mentioned proposal.
Consequently, RGB stereo data sets can be employed for training.
Second, we insert a novel color agnostic transform in front of the CNN.
Third, as the goal is to conduct cross-spectral disparity estimation by means of RGB training datasets, the transfer from training to inference is discussed.
An overview of the novel CADE approach is given in Fig.~\ref{img:CADE-intro} which shows the proposed training and inference modifications.

So far, RGB data is commonly used to train CNNs for stereo matching problems.
Additionally, artificially generated image content is mainly used for training due to the lack of camera-captured sequences.
CADE overcomes two challenges, namely, generating suitable multispectral training data from RGB images and deriving a color agnostic representation.
The latter decreases the differences between synthesized and real multispectral images as well as artificially generated and camera-captured training data.
Thus, CADE allows to train high-quality cross-spectral disparity estimators by using widely available artificially generated RGB stereo data for multispectral synthesis.

\subsection{Spectral Band Synthesis}
\label{sec:CADE-syn}
To close the gap between mono-modal and cross-spectral disparity estimation, the acquisition chain is discussed first.
A spectral intensity image
\begin{align}
	\!\!\!\! I_{c}[\mn] \!=\!\! \int_{0}^{\infty} \!\!\!\!\!\! s_{m,n}(\lambda) o_{m,n}(\lambda) l_{m,n}(\lambda) f_{m,n,c}(\lambda) r_{m,n}(\lambda) \mathrm{\,d}\lambda\!
	\label{eq:intensity-image}
\end{align}
is recorded by summing up the collected photons for every wavelength $\lambda$ at every spatial pixel position $(\mn)$. In (\ref{eq:intensity-image}), light source $s_{m,n}(\lambda)$ illuminates an object $o_{m,n}(\lambda)$ which reflects light.
The reflected light passes through the camera lens $l_{m,n}(\lambda)$, a spectral filter $f_{m,n,c}(\lambda)$ and is then measured by the camera with its response $r_{m,n}(\lambda)$.
Variable $c$ denotes the color plane index, e.g., $c \in \{ \text{R}, \text{G}, \text{B} \}$ for color images, or $c \in \{ \mathrm{BP}_{400}, \mathrm{BP}_{900} \}$ for bandpass filters with a central-wavelength (CWL) and a bandwidth of $50$ nm, respectively.
As all variables in $(\ref{eq:intensity-image})$ are fixed in our multispectral camera setup except for the spectral filters, $f_c(\lambda)$ is directly linked to the spectral intensity image $I_{c}[\mn]$.
Accordingly, another filter can then be simulated by composing intensity images, so that CADE is able to derive suitable filter curves and intensity images from RGB stereo data.
A given RGB stereo training set
\begin{align}
	\mathcal{D} = \bigcup_{k = 0}^{K-1} \ \bigcup_{v \,\in\, \mathcal{V} } \ \bigcup_{c \,\in\, \mathcal{C}} I_{k,v,c}[\mn]
\end{align}
contains $k \in [0,\dots, K\mathrm{-}1]$ stereo images with viewpoints\linebreak $v \in \mathcal{V} = \{\mathrm{left},\mathrm{right}\}$ and color indices $c \in \mathcal{C} = \{ \mathrm{R},\!\mathrm{G},\!\mathrm{B} \}$.
A single color channel $c$ is accessed via $I_{k,v,c}[\mn]$ as all images in the set are processed independently by CADE. Moreover, all images are assumed to be in the range $[0,1]$ and of floating point accuracy.
As with~\cite{genser2020_CSDL}, the unaltered red, green, and blue intensity images
\begin{align}
	\!\!\!\! \hat{I}_\mathrm{B}[\mn] \!\!\!\:=\!\!\!\: I_\mathrm{B}[\mn] \mathrm{,\,} \hat{I}_\mathrm{G}[\mn] \!\!\!\:=\!\!\!\: I_\mathrm{G}[\mn] \mathrm{,\,} \hat{I}_\mathrm{R}[\mn] \!\!\!\:=\!\!\!\: I_\mathrm{R}[\mn]
\end{align}
are included in the cross-spectral training set with $(\hat{\cdot})$ denoting the spectral estimation.
Furthermore, the composed images
\begin{align}
	\hat{I}_\mathrm{BG}[\mn]  &= (r_0 I_\mathrm{B}[\mn] + r_1 I_\mathrm{G}[\mn]) \, / \, (r_0+r_1)\\
	\hat{I}_\mathrm{BR}[\mn]  &= (r_2 I_\mathrm{B}[\mn] + r_3 I_\mathrm{R}[\mn]) \, / \, (r_2+r_3)\\
	\hat{I}_\mathrm{GR}[\mn]  &= (r_4 I_\mathrm{G}[\mn] + r_5 I_\mathrm{R}[\mn]) \, / \, (r_4+r_5)\\
	\hat{I}_\mathrm{BGR}[\mn] &= \dfrac{r_6 I_\mathrm{B}[\mn] + r_7 I_\mathrm{G}[\mn] + r_8 I_\mathrm{R}[\mn]}{r_6+r_7+r_8}
\end{align}
are added which corresponds to a weighted summation of spectral filter curves.
In above equation, uniformly distributed random parameters $r_0$ to $r_8$ are used to obtain arbitrary filter estimations $\hat{f}_c(\lambda)$.
While coefficients have been shared among equations in~\cite{genser2020_CSDL}, experiments revealed that an increase of randomness further improves cross-spectral training.
In contrast to~\cite{genser2020_CSDL}, we additionally introduce the combined images
\begin{align}
	\hat{I}_\mathrm{B \cap G}[\mn] \!&=\! \min\! \left( r_{9}  I_\mathrm{B}[\mn], r_{10} I_\mathrm{G}[\mn] \right) \mathrm{, \ } \forall (\mn)\\
	\hat{I}_\mathrm{G \cap R}[\mn] \!&=\! \min\! \left( r_{11} I_\mathrm{G}[\mn], r_{12} I_\mathrm{R}[\mn] \right) \mathrm{, \ } \forall (\mn)\\
	\hat{I}_\mathrm{B \cup G}[\mn] \!&=\! \max\! \left( r_{13} I_\mathrm{B}[\mn], r_{14} I_\mathrm{G}[\mn] \right) \mathrm{, \ } \forall (\mn)\\
	\hat{I}_\mathrm{G \cup R}[\mn] \!&=\! \max\! \left( r_{15} I_\mathrm{G}[\mn], r_{16} I_\mathrm{R}[\mn] \right) \mathrm{, \ } \forall (\mn)
\end{align}
making use of the pixel-wise minimum and maximum operators as well as the uniformly distributed random parameters $r_9$ to $r_{16}$.
These estimations roughly approximate an intersection or union of given filter curves.
For above equations, only pixel values that are actually contained in the recordings are used which avoids synthesis artifacts.
For training a CNN, color component $c$ can be chosen from the set
\begin{align}
	\hspace*{-0.6em}\hat{\mathcal{C}} \!\!\!\:=\!\!\!\: \{ &\mathrm{R},\! \mathrm{G},\! \mathrm{B},\! \mathrm{BG},\! \mathrm{BR},\! \mathrm{GR},\! \mathrm{BGR},\! \mathrm{B \!\cap\! G},\! \mathrm{G \!\cap\! R},\! \mathrm{B \!\cup\! G},\! \mathrm{G \!\cup\! R} \}\!\!
\end{align}
taking the eleven synthesized images $\hat{I}_c[\mn]$ into account.

\vskip-1em
\subsection{Color Agnostic Model}
\label{sec:CADE-ca}
CADE must cope with two major challenges: estimating synthesized images that are similar to real multispectral recordings and reducing the influence of artificially generated training data.
In practical scenarios, narrowband spectral filters are typically used which leads to photon shot and sensor electron noise.
Thus, images are filtered to obtain
\begin{align}
	\hat{I}^\mathrm{f}_c[\mn] = \median_{s \times s} \left( \hat{I}_c[\mn] \right)
\end{align}
using blocks of size $s \times s$ around the current pixel position.
For the centering, half the block size is denoted as $s_\mathrm{h} = \lfloor \nicefrac{s}{2} \rfloor$.
In the following, $s=3$ is chosen.
In contrast to~\cite{genser2020_CSDL}, which tries to simulate the noise during training, CADE uses a median filter to suppress noise while preserving the texture which is needed for disparity estimation.
This design enables to cope with a variety of camera sensors and spectral filters without needing to introduce prior assumptions.
Given the denoised spectral intensity images, a pixel-wise structural transform is applied to remove color information.
Thus, the average values
\begin{align}
	\mu_c[\mn] = \dfrac{1}{s^2} \sum_{i=-s_\mathrm{h}}^{s_\mathrm{h}} \sum_{j=-s_\mathrm{h}}^{s_\mathrm{h}} \hat{I}^\mathrm{f}_c[\mn]
\end{align}
and the variances
\begin{align}
	\sigma^2_c[\mn] = \dfrac{1}{s^2 - 1} \sum\limits_{i=\text{-}s_\mathrm{h}}^{s_\mathrm{h}} \sum\limits_{j=\text{-}s_\mathrm{h}}^{s_\mathrm{h}} \left( \hat{I}^\mathrm{f}_c[m\!+\!i,\!n\!+\!j] - \mu_c[m,\!n] \right)^{2}
\end{align}
are calculated for all pixel positions of the denoised image.
Given these local measures, the structural image
\begin{align}
	\hat{I}^\mathrm{s}_c[\mn] = \left( \hat{I}^\mathrm{f}_c[\mn] - \mu_c[\mn] \right) \, / \, \sigma_c[\mn] \mathrm{, \ } \forall (\mn)
\end{align}
is obtained.
However, as standard deviation $\sigma_c[\mn]$ can become very small or even zero, CADE uses the clipping
\begin{align}
	\hat{I}^\mathrm{a}_c[\mn] \!=\!
	\begin{cases}
		0, & \!\!\!\! \sigma_c[\mn] = 0\\
		\min\!\left(\! \max\!\left(\! \dfrac{1}{2} \!+\! \dfrac{\hat{I}^\mathrm{s}_c[\mn]}{2}, 1 \!\right)\!, 0 \!\right)\!, & \!\!\!\! \mathrm{else}\\
	\end{cases}\!\!\!\!\!\!\!\!\!\!
\end{align}
to obtain suitable color agnostic images $\hat{I}^\mathrm{a}_c[\mn]$ which now only contain the local image statistics of the recorded scene.
Since the neural networks require images in the range between 0 and 1, the image is shifted to the center of this range by 0.5.

\subsection{From Training to Inference}
\label{sec:CADE-sum}
Given the synthesized images $\hat{I}_c[\mn]$ with $c \!\in\! \hat{\mathcal{C}}$ from Sec.~\ref{sec:CADE-syn} and its color agnostic representation $\hat{I}^\mathrm{a}_c[\mn]$ from Sec.~\ref{sec:CADE-ca}, a set of spectral intensity training images
\begin{align}
	\mathcal{T} = \bigcup_{k = 0}^{K-1} \ \bigcup_{v \,\in\, \mathcal{V} } \ \bigcup_{c \, \in \, \hat{\mathcal{C}}} \hat{I}^\mathrm{a}_{k,v,c}[\mn]
\end{align}
is derived from the RGB stereo database $\mathcal{D}$.
Consequently, eleven training components result per color image which are used to train the CNN for cross-spectral stereo imaging.

The synthetically trained CNN can then be used to estimate disparities for camera-captured multispectral images.
In contrast to training, the stereo pairs are only preprocessed based on the color agnostic model of CADE for inference.

\section{Evaluation}
\label{sec:eval}
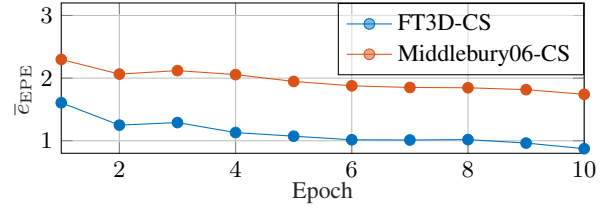
\begin{figure}[t]
	\small
	\setlength{\figurewidth}{0.85\columnwidth}
	\setlength{\figureheight}{2cm}
	\centering
%
%
\definecolor{mycolor1}{rgb}{0.00000,0.44700,0.74100}%
\definecolor{mycolor2}{rgb}{0.85000,0.32500,0.09800}%
\begin{tikzpicture}

\begin{axis}[%
width=0.951\figurewidth,
height=\figureheight,
at={(0\figurewidth,0\figureheight)},
scale only axis,
xmin=1,
xmax=10,
xlabel style={font=\color{white!15!black},below=-1.5mm},
xlabel={Epoch},
ymin=0.8,
ymax=3.2,
ylabel style={font=\color{white!15!black},below=1.5mm},
ylabel shift=12pt,
ylabel={$\overline{e}_\mathrm{EPE}$\strut},
axis background/.style={fill=white},
xmajorgrids,
ymajorgrids,
legend style={at={(1.0,1.0)},anchor=north east,legend cell align=left, fill=white, fill opacity=0.5, draw opacity=1, text opacity=1}
]
\addplot [color=mycolor1,mark=*]
  table[row sep=crcr]{%
1	1.60693836212158\\
2	1.24989342689514\\
3	1.29034895896912\\
4	1.13079011440277\\
5	1.07290998935699\\
6	1.01563470840454\\
7	1.01229162216187\\
8	1.01844263076782\\
9	0.963472671508789\\
10	0.873171830177307\\
};
\addlegendentry{FT3D-CS}

\addplot [color=mycolor2,mark=*]
  table[row sep=crcr]{%
1	2.29784\\
2	2.06419555555556\\
3	2.11968918032787\\
4	2.05627143631436\\
5	1.9470581342218\\
6	1.87693655650967\\
7	1.85078483082555\\
8	1.84586674713694\\
9	1.81613022993162\\
10	1.74144042747265\\
};
\addlegendentry{Middlebury06-CS}

\end{axis}
\end{tikzpicture}%
	\caption{Comparison of the average endpoint error $\overline{e}_\mathrm{EPE}$ for artificial and camera-captured images and CADE-GANet. The novel CADE approach achieves a good generalization of stereo matchers.}
	\label{img:artificial-camera-captured}
\end{figure}
As depicted in Fig.~\ref{img:CADE-intro}, the novel CADE algorithm is able to turn any stereo matching network into a cross-spectral disparity estimator.
Hence, the performance of CADE is analyzed for recently published CNNs, i.e., Guided Aggregation Net (GANet)~\cite{zhang2019GANet}, Hierarchical Discrete Distribution Decomposition Net (HD$^3$Net)~\cite{yin2019}, and Pyramid Stereo Matching Network (PSMNet)~\cite{chang2018}.
To give further anchor points, Census + SGM~\cite{zeglazi2017}, ZNCC + SGM~\cite{mattoccia2008}, and the recently published Dense Adaptive Self-Correlation (DASC) for multi-spectral correspondence estimation~\cite{kim2017DASC} are taken into consideration.
Moreover, CADE is also compared to our previous work from~\cite{genser2020_CSDL}.
To train the CNNs, a spectral training set was derived from the \textit{FT3D} database~\cite{ft3d2016} by applying the novel CADE algorithm.
Then, GANet, HD$^3$Net, and PSMNet were trained based on their respective training procedure as well as the parameters according to the original papers.
The number of epochs was set to 10 for all CNNs. For comparing the various methods, the percentage of Bad Matched Pixels (BMP)~\cite{scharstein2001_BMP} with thresholds 3 and 5 as well as the End-Point Errors (EPE)~\cite{otte1994_EPE} are calculated which are the most common metrics for comparing stereo matchers.
In the following, two different dataset labels are in use.
The first acronym "CS" indicates that the stereo database was decomposed into its color components for obtaining six cross-spectral registration tasks per image, i.e., R\textrightarrow G, R\textrightarrow B, G\textrightarrow R, G\textrightarrow B, B\textrightarrow R, B\textrightarrow G.
The second acronym "RGB" means that the disparity of color stereo images is calculated.
Accordingly, the RGB stereo pairs are decomposed into their color components resulting in the mono-modal registration tasks R\textrightarrow R, G\textrightarrow G, B\textrightarrow B.
After matching each stereo component, the disparity map is obtained by median filtering the three candidates per pixel.
Thus, each cross-spectral stereo matcher can also be evaluated for mono-modal disparity estimation.

In the first analysis, the performance of CADE is investigated for ten epochs of the novel CADE-GANet when processing the artificially generated \textit{FT3D-CS}~\cite{ft3d2016} and the camera-captured \textit{Middlebury06-CS}~\cite{middlebury2006} datasets.
CADE-GANet was chosen as it achieves the best performance among all investigated CNNs. The performance is evaluated by using the mean EPE score $\overline{e}_\mathrm{EPE}$. From Fig.~\ref{img:artificial-camera-captured}, it can be concluded that artificially generated images and the novel CADE training lead to a good generalization on camera-captured images, too.

In a second analysis, the performance of CADE is investigated for the artificially generated \textit{MonkaaFamily}~\cite{ft3d2016} stereo set, and the two camera-captured databases \textit{Middlebury05}~\cite{middlebury2005} and \textit{Middlebury14}~\cite{middlebury2014}.
Various state-of-the-art approaches and the novel CADE CNNs are investigated for the RGB and CS mapping.
Having a closer look at Tab.~\ref{table:eval-cade}, several conclusions can be drawn.
First, CADE-GANet is the best stereo-matcher for all datasets as well as color and cross-spectral stereo matching.
Second, CADE improves the performance of every CNN and outperforms the recently published CSDL training approach from~\cite{genser2020_CSDL}.
Third, the pristine CNNs trained by the authors achieve in general a superior quality compared to the classical approaches~\cite{zeglazi2017,mattoccia2008,kim2017DASC}.
However, the classical approaches always obtain a certain robustness for all datasets as well as the RGB and the CS mappings, while the pristine CNNs may fail if the training is not similar enough to the actual inference.
Fourth, CADE overcomes this limitation and significantly increases the robustness of all CNNs on all databases for every mapping.
Fifth, the CNNs trained by CADE achieve also the best quality for mono-modal stereo matching (RGB), which can be explained as CADE also decreases other stereo distortions, e.g., illumination differences.
Sixth, for the best state-of-the-art method~\cite{zhang2019GANet} and our novel CADE-GANet, the average EPE is reduced by $100\,\% - \tfrac{1.29\,+\,2.71\,+\,1.28}{1.54\,+\,2.86\,+\,2.36}\,\% \approx 22\,\%$ for the RGB and by $100\,\% - \tfrac{4.27\,+\,3.28\,+\,1.87}{5.09\,+\,5.77\,+\,5.07}\,\% \approx 41\,\%$ for the CS mapping averaged over the investigated datasets (see Tab.~\ref{table:eval-cade}).
Summarized, CADE significantly improves mono-modal as well as cross-spectral stereo imaging.
\begin{table}[t]
	\centering
	\footnotesize
	\tabulinesep=0.1em
	\setlength\tabcolsep{1.5pt}
	\caption{Comparison of various state-of-the-art mono-modal and cross-spectral stereo matchers on different datasets.}
	\begin{tabu} to \columnwidth {|X[l]||r|r|r||r|r|r|}\cline{2-7}
		\multicolumn{1}{c|}{}                                 & \multicolumn{3}{c||}{\textbf{Dataset RGB}} & \multicolumn{3}{c|}{\textbf{Dataset CS}}\\
		\hhline{~======}
		\multicolumn{1}{c|}{}                                 & \multicolumn{6}{c|}{\textbf{\textit{MonkaaFamily}~\cite{ft3d2016}}}\\\hline
		\textbf{Algorithm}                                    & $\overline{e}_{\mathrm{EPE}}$ & $\overline{e}_\mathrm{BMP3}$ & $\overline{e}_\mathrm{BMP5}$ & $\overline{e}_{\mathrm{EPE}}$ & $\overline{e}_\mathrm{BMP3}$ & $\overline{e}_\mathrm{BMP5}$\\\hline
		\textbf{Census + SGM~\cite{zeglazi2017}}              &  6.99                         & 32.5\,\%                     & 25.3\,\%                      & 17.15 & 44.1\,\% & 38.3\,\% \\
		\textbf{DASC~\cite{kim2017DASC}}                      &  8.02                         & 36.1\,\%                     & 27.9\,\%                      & 14.92 & 36.7\,\% & 32.5\,\% \\
		\textbf{ZNCC + SGM~\cite{mattoccia2008}}              & 10.72                         & 38.4\,\%                     & 31.3\,\%                      & 20.22 & 40.1\,\% & 35.9\,\% \\\hline
		\textbf{GANet~\cite{zhang2019GANet}}                  & 1.54                          & 7.1\,\%                      & 4.3\,\%                       & 5.09 & 32.8\,\% & 25.2\,\% \\
		\textbf{HD$^\mathbf{3}$Net~\cite{yin2019}}                        & 42.36                         & 44.0\,\%                     & 42.4\,\%                      &  82.98 & 50.0\,\% & 49.8\,\% \\
		\textbf{PSMNet~\cite{chang2018}}                      &  6.90                         & 58.9\,\%                     & 46.6\,\%                      & 10.07 & 65.8\,\% & 54.1\,\% \\\hline
		\textbf{CSDL-GANet~\cite{genser2020_CSDL}}            &  4.67                         & 16.9\,\%                     & 11.9\,\%                      & 12.42  & 30.6\,\% & 24.9\,\% \\
		\textbf{CSDL-HD$^\mathbf{3}$Net~\cite{genser2020_CSDL}}           &  6.21                         & 30.8\,\%                     & 26.9\,\%                      & 13.38 & 39.7\,\% & 35.0\,\% \\
		\textbf{CSDL-PSMNet~\cite{genser2020_CSDL}}           &  6.87                         & 30.4\,\%                     & 23.5\,\%                      & 10.49 & 36.9\,\% & 30.3\,\% \\\hline
		\textbf{Novel CADE-GANet}                             & \textbf{1.29}                 & \textbf{6.6\,\%}             & \textbf{4.2\,\%}              & \textbf{4.27} & \textbf{23.3\,\%} & \textbf{17.7\,\%} \\
		\textbf{Novel CADE-HD$^\mathbf{3}$Net}                &  5.59                         & 25.2\,\%                     & 22.3\,\%                      & 10.11 & 33.6\,\% & 30.6\,\% \\
		\textbf{Novel CADE-PSMNet}                            &  2.37                         & 15.1\,\%                     &  9.7\,\%                      & 5.55 & 24.9\,\% & 20.5\,\% \\
		\cline{1-1}\hhline{~======}
		\multicolumn{1}{c|}{}                                 & \multicolumn{6}{c|}{\textbf{\textit{Middlebury05}~\cite{middlebury2005} }}\\\hline
		\textbf{Census + SGM~\cite{zeglazi2017}}              & 12.96                         & 30.6\,\%                     & 25.9\,\%                      & 17.15 & 44.1\,\% & 38.3\,\% \\
		\textbf{DASC~\cite{kim2017DASC}}                      & 16.42                         & 29.8\,\%                     & 26.2\,\%                      & 24.92 & 36.1\,\% & 32.5\,\% \\
		\textbf{ZNCC + SGM~\cite{mattoccia2008}}              & 16.92                         & 29.6\,\%                     & 26.0\,\%                      & 20.22 & 40.1\,\% & 35.9\,\% \\\hline
		\textbf{GANet~\cite{zhang2019GANet}}                  & 2.86                          & 10.6\,\%                     &  7.4\,\%                      &  5.77 & 24.2\,\% & 18.0\,\% \\
		\textbf{HD$^\mathbf{3}$Net~\cite{yin2019}}                        & 7.50                          & 21.5\,\%                     & 19.7\,\%                      & 11.43 & 32.9\,\% & 28.8\,\% \\
		\textbf{PSMNet~\cite{chang2018}}                      & 18.35                         & 38.7\,\%                     & 37.4\,\%                      & 18.81 & 48.4\,\% & 47.1\,\% \\\hline
		\textbf{CSDL-GANet~\cite{genser2020_CSDL}}            &  9.43                         & 14.9\,\%                     & 12.3\,\%                      &  9.89 &  17.9\,\% &  14.5\,\% \\
		\textbf{CSDL-HD$^\mathbf{3}$Net~\cite{genser2020_CSDL}}           &  8.37                         & 24.3\,\%                     & 20.1\,\%                      & 12.4 & 35.1\,\% & 30.3\,\% \\
		\textbf{CSDL-PSMNet~\cite{genser2020_CSDL}}           &  6.51                         & 17.9\,\%                     & 14.9\,\%                      &  7.82 & 22.6\,\% & 18.4\,\% \\\hline
		\textbf{Novel CADE-GANet}                             & \textbf{2.71}                 & \textbf{8.3\,\%}             & \textbf{6.5\,\%}              & \textbf{3.28} & \textbf{10.6\,\%} & \textbf{7.7\,\%} \\
		\textbf{Novel CADE-HD$^\mathbf{3}$Net}                &  7.44                         & 21.4\,\%                     & 18.8\,\%                      & 9.5 & 27.8\,\% & 23.5\,\% \\
		\textbf{Novel CADE-PSMNet}                            &  3.29                         & 10.1\,\%                     &  7.6\,\%                      &  3.68 & 12.5\,\% & 9.3\,\% \\
		\cline{1-1}\hhline{~======}
		\multicolumn{1}{c|}{}                                 & \multicolumn{6}{c|}{\textbf{\textit{Middlebury14}~\cite{middlebury2014} }}\\\hline
		\textbf{Census + SGM~\cite{zeglazi2017}}              &  7.11                         & 33.3\,\%                     & 28.2\,\%                      & 11.01 & 46.7\,\% & 37.8\,\% \\
		\textbf{DASC~\cite{kim2017DASC}}                      &  7.37                         & 33.9\,\%                     & 32.3\,\%                      & 13.97 & 30.9\,\% & 27.5\,\% \\
		\textbf{ZNCC + SGM~\cite{mattoccia2008}}              &  7.10                         & 32.9\,\%                     & 28.4\,\%                      & 10.24 & 42.0\,\% & 35.3\,\% \\\hline
		\textbf{GANet~\cite{zhang2019GANet}}                  &  2.36                         & 12.5\,\%                     &  9.0\,\%                      &  5.07 & 22.0\,\% & 16.9\,\% \\
		\textbf{HD$^\mathbf{3}$Net~\cite{yin2019}}                        & 23.51                         & 40.1\,\%                     & 36.4\,\%                      & 59.38 & 57.7\,\% & 55.7\,\% \\
		\textbf{PSMNet~\cite{chang2018}}                      & 10.51                         & 46.1\,\%                     & 39.1\,\%                      & 14.51 & 46.6\,\% & 43.3\,\% \\\hline
		\textbf{CSDL-GANet~\cite{genser2020_CSDL}}            &  4.30                         & 16.8\,\%                     & 11.8\,\%                      &  5.01 &  21.2\,\% &  16.9\,\% \\
		\textbf{CSDL-HD$^\mathbf{3}$Net~\cite{genser2020_CSDL}}           &  5.73                         & 21.5\,\%                     & 16.4\,\%                      &  9.76 & 30.2\,\% & 26.8\,\% \\
		\textbf{CSDL-PSMNet~\cite{genser2020_CSDL}}           &  7.19                         & 22.9\,\%                     & 19.7\,\%                      &  8.50 & 31.8\,\% & 27.6\,\% \\\hline
		\textbf{Novel CADE-GANet}                             & \textbf{1.28}                 & \textbf{6.3\,\%}             & \textbf{4.1\,\%}              & \textbf{1.87} & \textbf{8.7\,\%} & \textbf{6.4\,\%} \\
		\textbf{Novel CADE-HD$^\mathbf{3}$Net}                &  3.49                         & 19.7\,\%                     & 15.5\,\%                      &  4.77 & 25.4\,\% & 19.8\,\% \\
		\textbf{Novel CADE-PSMNet}                            &  1.86                         &  7.5\,\%                     &  5.0\,\%                      &  2.63 & 14.0\,\% &  8.8\,\% \\
		\hline
	\end{tabu}
	\label{table:eval-cade}
	\vskip-1em
\end{table}
\par
\begin{figure}[t]
	\centering
	\begin{tikzpicture}
		\node[above right] (img) at (0,0) {\includegraphics[width=0.479\columnwidth]{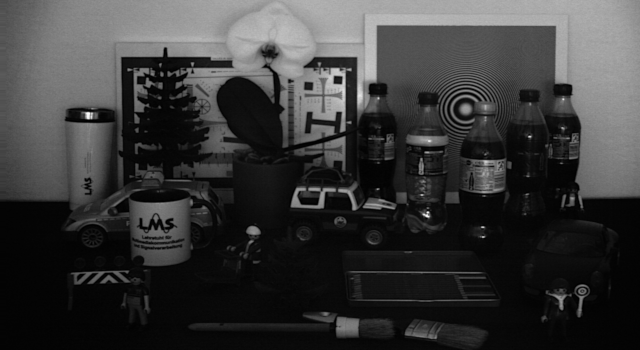}};
		\node[above right,fill=white,fill opacity=0.5,text opacity=1] at (3.37,2.01) {\scriptsize $\mathrm{BP}_{400}$};
	\end{tikzpicture}
	\hspace*{-0.9em}
	\begin{tikzpicture}
		\node[above right] (img) at (0,0) {\includegraphics[width=0.479\columnwidth]{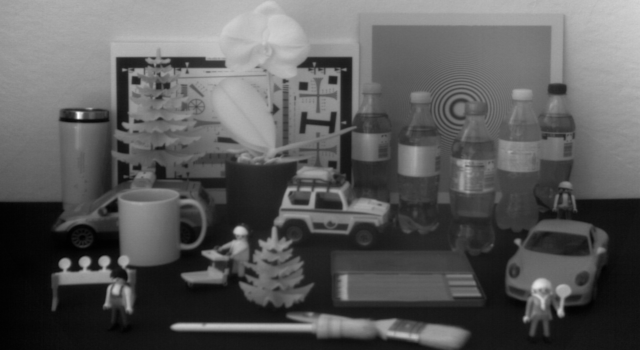}};
		\node[above right,fill=white,fill opacity=0.5,text opacity=1] at (3.37,2.01) {\scriptsize $\mathrm{BP}_{900}$};
	\end{tikzpicture}
	\vskip-0.3em
	\begin{tikzpicture}
		\node[above right] (img) at (0,0) {\includegraphics[width=0.479\columnwidth]{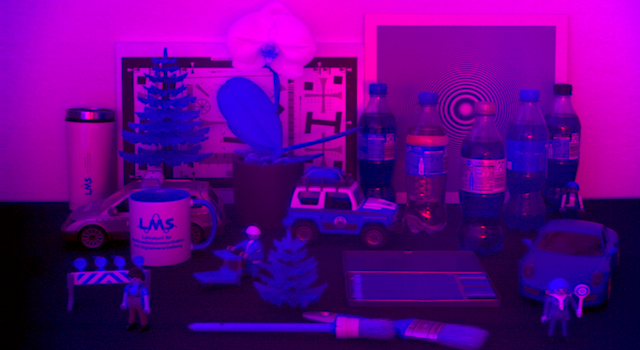}};
		\node[above right,fill=white,fill opacity=0.5,text opacity=1] at (1.15,1.99) {\scriptsize Registered: $\mathrm{BP}_{900}\!\!\rightarrow\!\!\mathrm{BP}_{400}$};
	\end{tikzpicture}
	\hspace*{-0.9em}
	\begin{tikzpicture}
		\node[above right] (img) at (0,0) {\includegraphics[width=0.479\columnwidth]{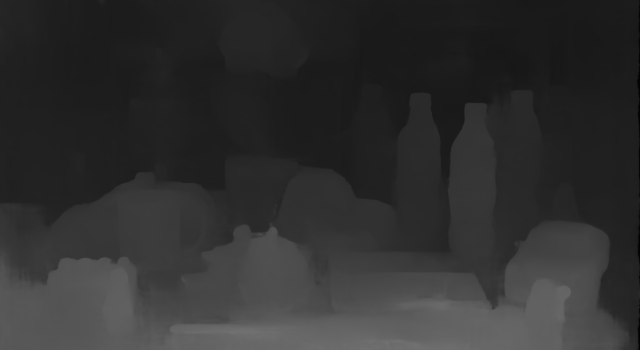}};
		\node[above right,fill=white,fill opacity=0.5,text opacity=1] at (2.71,1.99) {\scriptsize Disparity map};
	\end{tikzpicture}
	\vskip-0.75em
	\caption{Sample scene recorded using a self-manufactured multispectral stereo camera. In the top line, the $\mathrm{BP}_{400}$ (UV) and $\mathrm{BP}_{900}$ (NIR) images of a sample scene are depicted. In the second line, the registered composition of both images is depicted together with the estimated disparity map using the novel CADE-GANet. Best to be viewed enlarged.}
	\label{img:visual-example}
\end{figure}
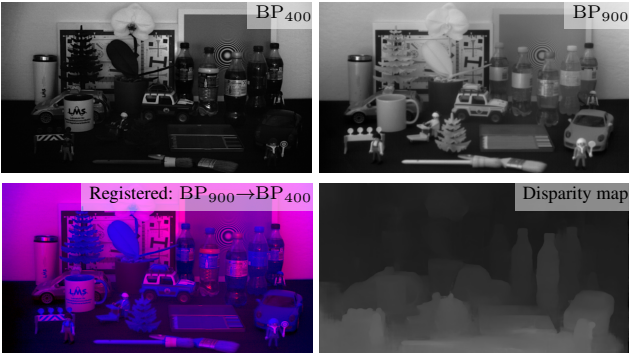

Fig.~\ref{img:visual-example} shows a scene recorded using a bandpass with CWL of 400 nm (UV) as well as a bandpass with CWL of 900 nm (NIR) stereo pair.
Obviously, the image content of the two intensity images is extremely diverse due to the large spectral distance.
However, CADE is still able to generate suitable disparity maps even for challenging multispectral registration tasks.
The occluded pixels of the registered image from Fig.~\ref{img:visual-example} can be reconstructed using cross spectral reconstruction~\cite{genser2020_CASE, sippel_deepguided_2023}.

\vfill
\section{Conclusion}
\label{sec:conclusion}
In this paper, a novel color agnostic disparity estimation is introduced to transform any stereo matching CNN into a cross-spectral disparity estimator.
The challenge of missing multispectral training data is circumvented by introducing a spectral synthesis in combination with a color agnostic transform, so that any available RGB stereo database can be used for cross-spectral disparity estimation.
Moreover, it is shown that the novel method generalizes well from artificial training to camera-captured inference.
In total, the average end-point error is decreased by $41$\,\% for cross-spectral and by $22$\,\% for RGB images compared to the state of the art.

\vfill\pagebreak\clearpage
\bibliographystyle{IEEEbib}
\bibliography{refs}

\end{document}